\documentclass{PoS}
\usepackage{epsfig}
\PoS{PoS(LAT2005)331}

\newcommand{\beq}{\begin{align*}}
\newcommand{\eeq}{\end{align*}}
\newcommand{\nn}{\nonumber}

\newcommand{\dd}{\partial}

\newcommand{\sbra}[1] { \left( #1 \right)}     

\title{
\begin{picture}(0,0)(0,0)%
   \put(350,75){\makebox(0,0)[l]{\textnormal{\normalsize CHIBA-EP-154}}}%
\end{picture}%
Understanding the U(1) problem through dyon configuration in the Abelian projection }

\ShortTitle{Understanding the U(1) problem through dyon configuration in the Abelian projection }

\author{\speaker{Seikou Kato}\\
        Takamatsu National College of Technology, Takamatu City 761-8058, Japan\\
        E-mail: \email{kato@takamatsu-nct.ac.jp}}

\author{Kei-Ichi Kondo\\
        Department of Physics, Faculty of Science, 
Chiba University, Chiba 263-8522, Japan\\
        E-mail: \email{kondok@faculty.chiba-u.jp}}

\abstract{
We give a short review of the recently obtained result that the magnetic monopole promoted to the dyon 
due to the vacuum angle $\theta$ resolves the U(1) problem in the sense that the dyon obtained in this 
way gives a dominant contribution to the topological susceptibility \cite{Kato-Kondo-05}. 
}

\FullConference{XXIIIrd International Symposium on Lattice Field Theory\\
25-30 July 2005\\
Trinity College, Dublin, Ireland}

\begin{document}

\section{Introduction}

It is very interesting and challenging to study a mechanism of the non-perturvative phenomena of QCD,
such as quark confinement, chiral symmetry breaking, strong CP violation, etc.
These non-perturbative phenomena are believed to be well understood in the unified way by considering 
the topologically nontrivial configurations of the gluon field.  
The U(1) problem or $\eta$ meson problem \cite{Weinberg75} is also one of such phenomena.
't Hooft \cite{tHooft76} pointed out that topologically nontrivial configurations such as instantons 
give the nonzero anomaly 
and suggested that instantons are the relevant topological objects related to the resolution of the 
U(1) problem\cite{instantonGroup}. 
However, it was not clear how to  compute the $\eta'$ mass. 
Moreover, it was pointed out that the Ward-Takahashi identity for the U$_A$(1) current with the anomalous 
term contradicts with the quark--antiquark condensation in the instanton $\theta$ vacuum \cite{Crewther77}.

\begin{figure}[h]
\vspace*{5mm}

\centerline{
\epsfxsize=0.8\textwidth
\epsfbox{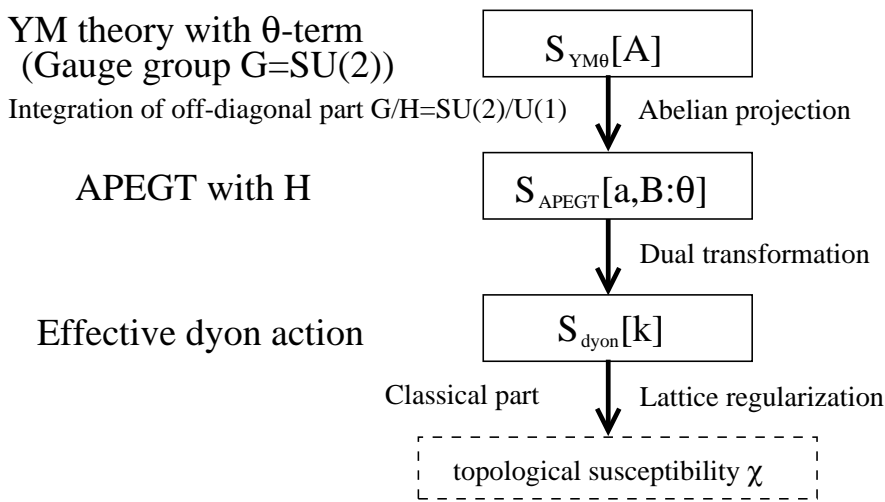}
}
\caption{A flowchart of our strategy}
\label{fig:strategy}
\end{figure}

In this talk, we review our recent result that the U(1) problem is understood through the
dyon configuration. 
A strategy for solving the U(1) problem along this line has already been discussed by Ezawa 
and Iwazaki~\cite{EI82} based on the idea of the Abelian projection proposed by 't~Hooft~\cite{tHooft81}. 
However, they assumed   in their analyses the {\it Abelian dominance} from the beginning and used  
an Abelian-projected effective theory which is conjectured to be derived from the Yang-Mills theory 
in the long distance.  
In contrast, extending the method developped by one of author\cite{Kondo97},
{\it we derive the Abelian-projected effective theory} based on the functional 
integration of the off-diagonal degrees of freedom from the Yang-Mills theory with the $\theta$ angle. 
We summarize a flowchart of our strategy in Figure \ref{fig:strategy}.

\section{Derivation of effective dyon action}

We start from the SU(2) Yang-Mills (YM) action $S_{YM}[A]$ 
with the $\theta$ term $S_{\theta}[A]$:
\begin{eqnarray}
S_{YM\theta}[A] &=&
S_{YM}[A]+S_{\theta}[A] ,
\label{model-1a}\\
S_{YM}[A] &=& -\frac{1}{2g^2}\int d^4x  {\rm tr}(F_{\mu\nu}F^{\mu\nu}) ,
\label{model-1b}\\
S_{\theta}[A] &=& \frac{\theta}{16\pi^2}
\int d^4x  {\rm tr}(F_{\mu\nu}\tilde{F}^{\mu\nu}) ,
\label{model-1c}
\end{eqnarray}

where the field strength $F_{\mu\nu}$ is defined by
\begin{eqnarray} 
F_{\mu\nu}(x) = \sum_{A=1}^{3} F^{A}_{\mu\nu}(x)T^{A} 
= \dd_{\mu} A_{\nu}(x) - \dd_{\nu} A_{\mu}(x) -ig[A_{\mu}(x),A_{\nu}(x)] ,
\label{model-1d1}
\end{eqnarray}
and $T^A(A=1,2,3)$ is the generator of the Lie algebra of the gauge group $SU(2)$.  
The Hodge dual $\tilde{F}_{\mu\nu}$ of $F_{\mu\nu}$  is defined by 
\begin{eqnarray} 
\tilde{F}_{\mu\nu}(x) \equiv 
\frac{1}{2}\epsilon_{\mu\nu\alpha\beta}F^{\alpha\beta}(x) .
\label{model-1d1-1}
\end{eqnarray}

\subsection{Step 1: Cartan decomposition}
We decompose $A_{\mu}$ into the diagonal U(1)  and the off-diagonal SU(2)/U(1) parts as
\begin{eqnarray} 
A_{\mu}(x)=a_{\mu}(x)T^3+ \sum_{a=1}^{2}A_{\mu}^a(x)T^a \in H \oplus (G-H).
\label{model-1d}
\end{eqnarray}
where $a_{\mu}(x)$ and $A_{\mu}^a(x)$ are diagonal, off-diagonal gluon field, respectively.
Accordingly, the field strength $F_{\mu\nu}$ is decomposed as 
\begin{eqnarray} 
F_{\mu\nu} &=& 
[f_{\mu\nu}(x)+C_{\mu\nu}(x)]T^3 + S_{\mu\nu}^a(x)T^a, \\
f_{\mu\nu}(x) &\equiv& \dd_{\mu}a_{\nu}(x)-\dd_{\nu}a_{\mu}(x), \\
S_{\mu\nu}^a(x) &\equiv& D_{\mu}[a]^{ab}A_{\nu}^b(x)
-D_{\nu}[a]^{ab}A_{\mu}^b(x),\\
C_{\mu\nu}(x)T^3 &\equiv& -i[A_{\mu}(x),A_{\nu}(x)] ,
\label{model-2}
\end{eqnarray}
where the covariant derivative $D_{\mu}[a]$ is defined by
\begin{eqnarray} 
D_{\mu}[a] =\dd_{\mu}+i[a_{\mu}T^3,\cdot] , \quad
D_{\mu}[a]^{ab} =\dd_{\mu}\delta^{ab}-\epsilon^{ab3}a_{\mu} .
\label{model-3}
\end{eqnarray}

Next, the total action $S_{YM\theta}[A]$ is decomposed as
\begin{eqnarray} 
S_{YM\theta}[A] &=& \int d^4x
\left\{
-\frac{1}{4g^2}f_{\mu\nu}f^{\mu\nu}
+\frac{\theta}{32\pi^2}f_{\mu\nu}\tilde{f}^{\mu\nu}
-\frac{1}{4}g^2B_{\mu\nu}B^{\mu\nu}
+\frac{1}{2g^2}A^{\mu}{}^aQ_{\mu\nu}^{ab}A^{\nu}{}^b 
\right\}  ,
\label{model-16a}
\\
Q_{\mu\nu}^{ab} &\equiv&  (D_{\rho}[a]D_{\rho}[a])^{ab}\delta_{\mu\nu}
-2\epsilon^{ab3}f_{\mu\nu} +g^2c_1\epsilon^{ab3}\tilde{B}_{\mu\nu}
+g^2c_0\epsilon^{ab3}B_{\mu\nu} \nn\\
&& -D_{\mu}[a]^{ac}D_{\nu}[a]^{cb},
\label{model-16b}
\end{eqnarray}
where we have introduced the auxiliary (antisymmetric tensor) field $B_{\mu\nu}$ according to 
\begin{eqnarray} 
B_{\mu\nu}=g^{-2}(c_0 C_{\mu\nu}+c_1 \tilde{C}_{\mu\nu}),
\label{model-7b}
\end{eqnarray}
and two coefficients, $c_0$ and $c_1$ are determined as
\begin{eqnarray} 
c_0 = \sqrt{\frac{1}{2}
\sbra{-1+\sqrt{1+\sbra{\frac{g^2\theta}{8\pi^2}}^2}}}
,\quad
c_1 = \sqrt{\frac{1}{2}
\sbra{1+\sqrt{1+\sbra{\frac{g^2\theta}{8\pi^2}}^2}}} 
\label{model-13}
\end{eqnarray}
to be equivalent to the original action $S_{YM\theta}[A]$.

\subsection{Step 2: Maximaly abelian gauge fixing}

We perform the abelian projection in terms of BRST formalism and
consider a gauge fixing for the off-diagonal part;
\begin{eqnarray} 
F^{\pm}[A,a] \equiv (\dd^{\mu}\pm i\xi a^{\mu})A_{\mu}^{\pm}=0 ,
\label{gf1}
\end{eqnarray}
where we have used the $(\pm,3)$ basis,
$
O^{\pm} \equiv (O^1 \pm i O^2)/\sqrt{2} .
$
\ Here the gauge parameter $\xi=0$ corresponds to the Lorentz gauge and $\xi=1$ to 
(the differential form of) the maximal abelian gauge (MAG). 
In the BRST quantization, the GF condition (\ref{gf1}) amounts to adding the following GF term 
and the Faddeev--Popov (FP) term \cite{Kondo97},
\begin{eqnarray} 
{\cal L}_{GF+FP} = \phi^aF^a[A,a]+\frac{\alpha}{2}(\phi^a)^2 
+i\bar{c}^aD^{\mu ab}[a]^{\xi}D_{\mu}^{bc}[a]c^c 
-i\xi \bar{c}^a[A_{\mu}^aA^{\mu b}-A_{\mu}^cA^{\mu c}\delta^{ab}]c^b ,
\label{gf3}
\end{eqnarray}
where $\phi$ stands for the Langange multiplier field and $F^a[A,a]$ is a 
gauge fixing function defined by
\begin{eqnarray} 
F^a[A,a]=(\dd^{\mu}\delta^{ab}-\xi\epsilon^{ab3}a^{\mu})A_{\mu}^b
= D^{\mu ab}[a]^{\xi}A_{\mu}^b .
\label{gf4}
\end{eqnarray}
Thus the total Lagrangian is obtained by adding (\ref{gf3}) to (\ref{model-16a}),
\begin{eqnarray} 
{\cal L} = {\cal L}_{YM\theta}[A] + {\cal L}_{GF+FP}.
\label{gf5}
\end{eqnarray}
We choose gauge parameters $\alpha=\xi=1$ below.

\subsection{Step 3: APEGT with $H$}

We integrate out the off-diagonal fields, $\phi^a$, $A_{\mu}^a$, $c^a$, $\bar{c}^a$ 
belonging to the Lie algebla of SU(2)/U(1) and obtain the Abelian-projected effective 
gauge theory (APEGT) written in terms of the diagonal fields, $a_{\mu}$ and $B_{\mu\nu}$
\footnote{See \cite{Kato-Kondo-05} for the detail of the calculations.}.
As a result, we obtain APEGT
\footnote{We have neglected the ghost self-interaction terms and higher 
derivative terms.};
\begin{eqnarray} 
S_{APEGT}[a,B:\theta] &=& \int d^4x \left[
-\frac{1+z_a-z'_a}{4g^2}f_{\mu\nu}f^{\mu\nu}
-\frac{1+z_b}{4}g^2B_{\mu\nu}B^{\mu\nu}
\right.\nonumber\\
&& \left.
-\frac{1}{2}z_cB_{\mu\nu}\tilde{f}^{\mu\nu} 
+\frac{\theta}{32\pi^2}f_{\mu\nu}\tilde{f}^{\mu\nu} 
+\frac{1}{2}z_df_{\mu\nu}B^{\mu\nu}
+\frac{1}{2}z_eB_{\mu\nu}\tilde{B}^{\mu\nu}
\right.\nonumber\\
&& \left.
+ (\mbox{4-ghost terms})
+(\mbox{higher derivative terms})
\right] ,
\label{lndet-11}
\end{eqnarray}
where $z$'s are renormalization constants; 
\begin{eqnarray} 
z_a &=& -\frac{10}{3}\kappa\frac{g^2}{16\pi^2}\ln\mu^2, \quad
z'_a = \frac{1}{3}\kappa\frac{g^2}{16\pi^2}\ln\mu^2, \quad
z_b = \kappa\frac{g^2}{16\pi^2}\ln\mu^2, \nonumber\\
z_c &=& 2\kappa c_1\frac{g^2}{16\pi^2}\ln\mu^2, \quad
z_d = 2\kappa c_0\frac{g^2}{16\pi^2}\ln\mu^2, \quad
z_e = -\kappa\frac{g^4}{16\pi^2}\cdot \frac{g^2\theta}{16\pi^2}\ln\mu^2.  
\label{lndet-9}
\end{eqnarray}
We have introduced the second Casimir operator $\kappa$ which is given for $G=SU(2)$ by 
\begin{eqnarray} 
\kappa\equiv C_2(G)=\epsilon^{3ab}\epsilon^{3ab}=2 .
\end{eqnarray}

\subsection{Step 4 : Effective dyon action}

The  magnetic monopole current $k_{\mu}$ is defined by 
\begin{eqnarray} 
k^{\mu}\equiv \dd_{\nu}\tilde{f}^{\mu\nu}, \quad 
\tilde{f}^{\mu\nu}=\frac{1}{2}\epsilon^{\mu\nu\rho\sigma}f_{\rho\sigma} .
\label{apegt-6}
\end{eqnarray}

Introducing the monopole currents in the APEGT (\ref{lndet-11}) and integrating out of 
all fields except for $k_{\mu}$, we obtain 
\begin{eqnarray} 
S_{dyon}[k] =  \int d^4x 
\frac{1}{2}g_m^2[\theta]k^{\mu}(x)D_{\mu\nu}k^{\nu}(x),
\label{apegt-17}
\end{eqnarray}
where $g_m[\theta]  := g |\tau|  = \sqrt{g_m^2 + q_m^2}$ 
($g_m \equiv 4\pi/g$,$\quad q_m \equiv g\theta/2\pi$)
and the kernel $D_{\mu\nu}$ stands for the massless vector propagator, e.g., 
$D_{\mu\nu}=(1/\partial^2)(\delta_{\mu\nu} - \partial_\mu \partial_\nu/\partial^2)$ in the Landau gauge. 

Note that the monopole current $k_{\mu}(x)$  acquires the electric charge and the magnetic 
monopole is changed 
to the dyon due to the existence of the $\theta$ term in agreement with the Witten effect 
\cite{Witten:1979ey2}.
{\it Thus, we completely reproduced the effective dyon action obtained in \cite{EI82} without 
any assumptions.}

\section{Topological susceptibility and Witten-Veneziano formula}

Let us argue that {\it the dyon configuration is the most relevant one
for solving the U(1) problem} in SU(2) QCD by evaluating the 
topological susceptibility from the dyon 
configuration appearing in the APEGT with $\theta$-term.  
To estimate the numerical value of the topological susceptibility, 
we consider the lattice regularized version of (\ref{apegt-17}), 
\begin{eqnarray} 
S_E = \sum_{x,y}\sbra{\bar{\beta}+\frac{\theta^2}{\bar{\beta}}}
k^{\mu}(x)D_{\mu\nu}(x-y)k^{\nu}(y), \quad
\bar{\beta} \equiv \frac{1}{2}\sbra{\frac{4\pi}{g}}^2.
\label{WV-2}
\end{eqnarray}
According to the analysis of the monopole action by the inverse Monte-Carlo simulation,  
the self-mass term of the monopole current is dominant in the low-energy region, e.g., 
$G_2/G_1\simeq 0.33$ at the scale $1.7 \rm{fm}$ where 
 $G_1$ and $G_2$ are respectively the self-coupling and the nearest-neighbor coupling of 
the monopole current \cite{Kato:2000}.
Furthermore, the monopole configuration subject to
$|k_{\mu}(x)|=1$ is dominant in the low-energy region \cite{SS91} and
the energy density $e_\theta$ is written as 
\begin{eqnarray} 
e_\theta = S_E/V \simeq \sbra{\bar{\beta}+\frac{\theta^2}{\bar{\beta}}}D(0) .
\label{WV-4}
\end{eqnarray}
Therefore, the topological susceptibility $\chi_E$ is calculated:
\begin{eqnarray} 
\chi_E \equiv \sbra{\frac{d^2 e_\theta}{d\theta^2}}_{\theta=0}
\simeq \frac{2}{\bar{\beta}}D(0) .
\label{WV-5}
\end{eqnarray}
The result of quantum perfect lattice action for monopole 
obtained by Chernodub et al.\cite{Kato:2000} show 
$\bar{\beta}D(0)\equiv G_1 =0.059$ and $\bar{\beta}=2.49$
at the physical scale $b=3.8\sigma_{phys}^{-1/2}$ . 
(Note that 
$b=1\sigma_{phys}^{-1/2}$ corresponds to $1.7 \rm{fm}$, provided that the string tension $\sigma_{phys} \cong (440\rm{MeV})^2$ in SU(2) QCD.
)
 By substituting these values into (\ref{WV-5}), the topological susceptibility is determined as
\begin{eqnarray} 
\chi_E^{1/4}/\sigma_{phys}^{1/2}=0.371 ,
\label{WV-6}
\end{eqnarray}
in units of the string tension $\sigma_{phys}$. 
Remarkably, this estimate reproduces $76\%$ of the full result 
\begin{eqnarray}
  \chi^{1/4}/\sigma_{phys}^{1/2}=0.486\pm 0.010 ,
\label{Teper}
\end{eqnarray}
obtained by  Teper \cite{Teper98} in the simulation of SU(2)QCD. 
Note that our result is also consistent with the mass formula for $\eta'$, the so-called the  
Witten-Veneziano formula\cite{Veneziano79}.

Thus we conclude  that {\it the dyon, i.e., magnetic monopole with the electric charge 
proportional to the vacuum angle $\theta$, gives dominant contribution to the topological 
susceptibility}.

\section{Conclusion}

In this paper, we have given a short review of how to 
derive the effective dyon action directly from SU(2) Yang-Mills theory with $\theta$-term
by using BRST formalism.
By estimating the classical part of the dyon action, we have calculated the topological 
susceptibility.  
The obtained value agrees with the numerical result obtained by the recent lattice gauge theory. 
Thus we have shown that the dyon-like configuration gives a dominant contribution to the 
topological susceptibility and resolves the U(1) problem.



\end{document}